\documentclass[preprint,aps,superscriptaddress,nofootinbib,showpacs]{revtex4-1}
\usepackage[dvipdfmx]{graphicx}
\usepackage{epsfig}
\usepackage{bm}
\usepackage{amssymb}
\usepackage{wrapfig}
\usepackage[usenames, dvipsnames]{color}

\newcommand{\cM}{{\cal M}}
\newcommand{\cO}{{\cal O}}

\newcommand{\psitld}{\mbox{${\tilde{\psi}}$}}

\newcommand{\vq}{\mbox{$\bm{q}$}}
\newcommand{\vbr}{\mbox{$\bm{r}$}}

\newcommand{\vB}{\mbox{$\bm{B}$}}

\begin{document}

\title{\bf
Pion Production via Proton Synchrotron Radiation \\
in Strong Magnetic Fields in Relativistic Field Theory:\\
Scaling Relations and Angular Distributions}

\author{Tomoyuki~Maruyama}
\affiliation{College of Bioresource Sciences,
Nihon University,
Fujisawa 252-8510, Japan}
\affiliation{Advanced Science Research Center,
Japan Atomic Energy Agency, Tokai 319-1195, Japan}
\affiliation{National Astronomical Observatory of Japan, 2-21-1 Osawa,
Mitaka, Tokyo 181-8588, Japan}

\author{Myung-Ki Cheoun}
\affiliation{Department of Physics, Soongsil University, Seoul,
156-743, Korea}
\affiliation{National Astronomical Observatory of Japan, 2-21-1 Osawa,
Mitaka, Tokyo 181-8588, Japan}

\author{Toshitaka~Kajino}
\affiliation{National Astronomical Observatory of Japan, 2-21-1 Osawa,
Mitaka, Tokyo 181-8588, Japan}
\affiliation{Department of Astronomy, Graduate School of Science,
University of Tokyo, Hongo 7-3-1, Bunkyo-ku, Tokyo 113-0033, Japan}

\author{Grant J. Mathews}
\affiliation{Center of Astrophysics, Department of Physics,
University of Notre Dame, Notre Dame, IN 46556, USA}

\date{\today}


\begin{abstract}
We study pion production by proton synchrotron radiation
in the presence of a strong magnetic field when the Landau numbers of 
the initial and final protons are $n_{i,f} \sim 10^4 - 10^5$.  
We find in our relativistic field theory calculations that the pion decay width
depends only on the field strength parameter which previously  was 
only conjectured based upon semi-classical arguments.
Moreover, we also find new results that 
the decay width satisfies a robust scaling relation,
and that the polar angular distribution  of emitted pion momenta
is very narrow and can be easily obtained.
This scaling implies that one can infer the decay width in more realistic
 magnetic fields of $10^{15}$G, where $n_{i,f} \sim 10^{12} - 10^{13}$,
from the results for $n_{i,f} \sim 10^4 - 10^5$.
The resultant pion intensity and angular distributions for realistic
 magnetic field strengths are presented and their physical implications
discussed. 
\end{abstract}

\maketitle



It is widely accepted that soft gamma repeaters (SGRs)
and anomalous X-ray pulsars (AXPs) correspond to magnetars
\cite{Mereghetti08}, and that the associate strong magnetic fields may have
a significant role in the production of high energy photons.
Furthermore, short duration gamma-ray bursts (GRBs) may arise from highly
magnetized neutron stars \cite{Soderberg06} or mergers
of binary neutron stars \cite{Gehrels05, Bloom026, Tanvit13},
and the most popular theoretical models for long-duration GRBs
invoke \cite{MacFadyen99, MacFadyen01, Harikae09, Harikae10}
 a magnetized accretion disk around neutron stars or rotating
black holes (collapsars) for their central engines.
Such magnetars (or black holes with strong magnetic fields)
have also been proposed \cite{Hillas84,Aarons03} as
an acceleration site for ultra high-energy (UHE) cosmic rays
(UHECRs) and a possible association \cite{Ioka05} between
magnetic flares and UHECRs has also been observed.

In this letter we consider the synchrotron emission that can be produced
by high-energy protons accelerated in such environments containing
a strong magnetic field.
This process has been proposed as a source for high-energy photons
in the GeV $-$ TeV range
\cite{Gupta07,Boettcher98,Totani98,Fragile04,Asano07,Asano09},
possibly in association with GRBs.
Of  interest to the present work is the fact that
 meson-nucleon coupling is in about 100 times larger than the
photon-nucleon coupling, and the meson production process is expected to
exceed photon synchrotron emission in the high energy regime.
For example, Refs. \cite{Ginzburg65a,Ginzburg65b,Zharkov65,TK99,BDK95}
addressed the possibility of $\pi^0$ emission from  protons in a strong
magnetic field.
The subsequent decay of such $\pi^0$s may be an additional source of
the observed TeV gamma rays in association with
supernova remnants \cite{Abdo-Sci10}.

However, previous calculations were performed in a semi-classical approximation
to the exact relativistic quantum-mechanical treatment of the proton
transitions among the Landau levels of the strong magnetic field.  Also,
there is ambiguity in the literature as to the proper behavior of
production in the strong field limit (cf. \cite{BDK95}).  
In this letter we resolve this ambiguity in an exact (Quantum Field Theory)
QFT calculation.

In our previous work \cite{P2Pi-1}
 we exploited the Green's function method for the propagation
of protons in a strong magnetic field,
and studied the pion production  from proton synchrotron emission
in a relativistic quantum approach,
whereby the pion is produced from the transition of a proton between
two Landau levels.

We then deduced the energy and angular distribution of emitted pions
which had not been deduced in previous semi-classical approaches.
Furthermore, we found that the anomalous magnetic moment (AMM) of the proton
 enhances the emission decay width by
about a factor of 50.
This huge effect comes from the fact that the overlap integral
of the two harmonic oscillator (HO) wave functions
is significantly altered by a small shift owing to the AMM
of the HO quantum numbers in the pion production energy region \cite{P2Pi-1}.

In that previous work, however, it was untractable to numerically evaluate
Landau numbers greater than $\sim 300$.
Therefore, we calculated the emission rate for a magnetic field strength
of $B \approx 5 \times 10^{18}$G for which fewer Landau levels were required.

In the present work we develop a new method and a new scaling relation to
give results for a much larger number of the Landau levels and for 
realistic magnetic field strengths.

In the semi-classical approach for the production of synchrotron radiation,
the magnetic field strength is characterized by the curvature parameter,
$\chi = e_i^3/(m^3 R_c)$
given in terms of  the incident particle mass $m$, its energy $e_i$,
and the curvature radius $R_c$.
For protons propagating in a strong magnetic field, the value of $\chi$
can be written as
\begin{equation}
\chi = {e_i^2 \over {m_p^3 R_c}} = {e B e_i \over {m_p^3}} ,
\end{equation}
where $m_p$ is the proton mass.
Pion production is the dominant process compared to
direct photon emission when $\chi \sim 0.01 - 1$ \cite{TK99}.

There are various semi-classical calculations
 \cite{Ginzburg65a,Ginzburg65b,Zharkov65,TK99,BDK95}
 which give different results,
but all of those models suggest that the decay width depends only
on the parameter $\chi$, not on the initial energy $e_i$
and the strength of the magnetic field $B$ individually,
though this has not been confirmed.

In this work, therefore, we have developed a new numerical method,
so that we can examine the scaling relation
for magnetic field strengths $\sim 10^{17} - 10^{18}$G,
that are weaker than on our work \cite{P2Pi-1}.
Though this strength is still large, we also find a new robust scaling rule
in the quantum calculation.
Based upon this we can realistically estimate for the first time
the results for a much  larger number of  Landau numbers,
i.e. lower magnetic field than previous calculations, 
from the results obtained from a smaller number of Landau levels.
%

Here, we briefly explain our approach.

We assume a uniform magnetic field along the $z$-direction,
$\vB = (0,0,B)$, and take the electro-magnetic vector potential $A^{\mu}$ to be
$A = (0, 0, x B, 0)$ at the position $\vbr \equiv (x, y, z)$ .

The relativistic proton wave function $\psitld$ is obtained
from the following Dirac equation:
\begin{equation}
\left[ \gamma_\mu \cdot (i \partial^\mu - e A^\mu) - m_p
- \frac{e \kappa_p}{2 m_p} \sigma_{\mu \nu}
(\partial^\mu A^\nu - \partial^\nu A^\mu ) \right]
\psitld (x) = 0 ,
\label{DirEq}
\end{equation}
where $\kappa_p$ is the proton AMM and $e$ is the elementary charge.
Here, we scale all variables with $\sqrt{eB}$ as
$X_\mu  = \sqrt{eB} x_\mu$ and $M_p = m_p /\sqrt{eB}$.
The proton single particle energy is then written as
\begin{eqnarray}
&& E(n, P_z, s) = \sqrt{ P_z^2 + (\sqrt{2n  + M_p^2}
- s \kappa_p/M_p)^2}.
\label{Esig}
\end{eqnarray}

When we use  the pseudo-vector coupling for the $\pi N$-interaction,
 we can obtain the differential decay width of the proton as
\begin{equation}
\frac{d^3 \Gamma_{p \pi} / \sqrt{eB}}{d Q^3} =
\frac{1}{8 \pi^2 E_\pi}  \left( \frac{f_\pi}{M_{\pi}} \right)^2
\sum_{n_f,s_f}   \frac{\delta(E_f + E_{\pi} - E_{i})}{4 E_i E_f} W_{if}~~,
\label{dfWid}
\end{equation}
with
\begin{equation}
W_{if} = {\rm Tr} \left\{ \rho_M (n_i, s_i, P_z) \cO_{\pi}
\rho_M (n_f, s_f, P_z - Q_z) \cO_{\pi}^{\dagger}
 \right\} ,
\end{equation}
where  $f_\pi$ is the pseudo-vector pion-nucleon coupling constant,
$M_\pi = m_{\pi} /\sqrt{eB}$ with $m_\pi$ being the pion mass,  and
\begin{eqnarray}
\rho_M &=&
\left[ E \gamma_0 + \sqrt{2n} \gamma^2 - P_z \gamma^3
 + M_p + (\kappa_p/M_p) \Sigma_z \right]
 \nonumber \\ && \quad \times
\left[ 1 + \frac{s}{\sqrt{ 2n + M_p^2} } \left(
 \kappa_p/M_p  + P_z \gamma_5 \gamma_0 - E \gamma_5 \gamma^3 \right)
 \right] ,
%
\\
\cO_{\pi}
&=&
\gamma_5 \left\{ \left[
\cM \left( n_i, n_f \right) \frac{1 + \Sigma_z}{2}
+  \cM \left( n_i - 1, n_f -1 \right) \frac{1 - \Sigma_z}{2} \right]
\left[ \gamma_0 Q_0 - \gamma^3 Q_z \right]
\right.\nonumber \\ &&  \quad \left.
- \left[ \cM \left( n_i, n_f-1 \right) \frac{1 + \Sigma_z}{2}
+  \cM \left( n_i - 1, n_f \right)
\frac{1 - \Sigma_z}{2} \right] \gamma^2 Q_T  \right\} .
\end{eqnarray}

In the above equation, the pion momentum scaled by $\sqrt{eB}$ is written as
$Q \equiv q / \sqrt{eB} =(E_{\pi}, 0, Q_T, Q_z)$, and
the HO overlap function $\cM(n_1, n_2)$ is defined \cite{P2Pi-1} as 
\begin{eqnarray}
 \cM (n_1,n_2)  & =&
\int d x f_{n_1}\left( x - \frac{Q_T}{2} \right)
f_{n_2} \left( x+ \frac{Q_T}{2} \right) .
\label{TrStM}
\end{eqnarray}

For these conditions the system is translationally  symmetric, and  quantities
with a finite $p_{iz}$ are given by a Lorentz transformation along
the $z$-direction.
For example, the decay width can be written as
$\Gamma_{p \pi} (p_z) =  \Gamma_{p \pi} (p_z=0) \sqrt{1 - (p_z/e_i)^2}$.
Then, we restrict the calculations to $p_{iz} = 0$ at first.

In the realistic condition, $B \sim 10^{15}$G and $\chi = 0.01 - 1$,
the initial Landau number becomes $n_i \sim 10^{12} - 10^{14}$, and
it is almost impossible to calculate $\cM(n_i, n_f)$ directly.  
In the following, we introduce an approximate, but very efficient, method for alleviating this  difficulty.

\begin{figure}[htb]
\begin{center}
{\includegraphics[scale=0.5,angle=270]{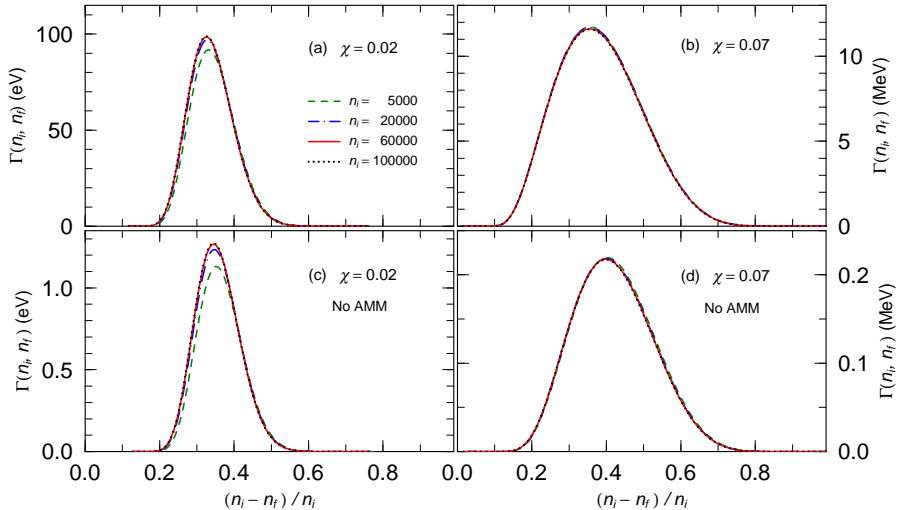}}
\caption{\small
(Color online) Pion decay widths of protons when $s_i = - s_f = -1$
as a function of $(n_i - n_f)/n_i$
for $\chi =0.02$ (a) and $\chi = 0.07$ (b). The bottom panels (c) and (d) show the results without the AMM
for $\chi = 0.02$ and $\chi = 0.07$, respectively.
The dashed, dotted-dashed solid and dotted lines
represent the results with initial Landau number,
$n_i = 5 \times 10^3$, $2  \times 10^4$, $6  \times 10^4$
and $10^5$, respectively.
}
\label{WidIFb}
\end{center}
\end{figure}

In Fig.~\ref{WidIFb} we present the decay widths as a function
of $(n_i - n_f)/n_i$
when the parameter $\chi$ is fixed.
The upper panels show the results with the AMM included for $\chi =0.02$ (a)
and $\chi = 0.07$ (b), and the lower panels show the results
without the AMM for $\chi =0.02$ (c) and $\chi = 0.07$ (d).
The dashed, dot-dashed,  solid and dotted lines
show the results with initial Landau numbers,
$n_i = 5 \times 10^3$, $2  \times 10^4$, $6  \times 10^4$
and $10^5$, respectively.
In all results the initial and final proton spins are set to be
$s_i = - s_f = -1$
because this is the dominant contribution.

First, we note that the dependence of the decay width on 
$\Delta n_{if}/n_i = (n_i - n_f)/n_i$
for the different cases is almost completely identical.  
That is, they nearly overlap when $\chi$ is fixed.
In particular, the difference between $n_i = 6 \times 10^4$ and $n_i =10^5$
is not discernible.
There are however various semi-classical calculations
 \cite{Ginzburg65a,Ginzburg65b,Zharkov65,TK99,BDK95}
 which give different results,
but all of the models indicate that the decay width depends only on the
parameter $\chi$, not on the initial energy $e_i$ and
the magnetic field $B$.
Thus, we confirm that this scaling relation is satisfied in
the quantum calculations, particularly for $n_i \gtrsim 6 \times 10^4$,
independently of the AMM.

In the all results the peak positions are at
$\Delta n_{if} / n_i \approx 0.3 - 0.4$.
As $\chi$ increases, the peak position is only slightly shifted to a larger
value, and the peak width becomes slightly broader.

When comparing the results with the AMM included (a,b) and
those without the AMM (c,d),
 we see that the AMM still increases the decay widths significantly
even in the present conditions where magnetic fields are much weaker
and the initial Landau  numbers are much larger  than those
in the previous work, $B=5 \times 10^{18}$G and $n_i \approx 48$.
\cite{P2Pi-1}.


Moreover, we see that the peak position is shifted by
including the AMM with the scaling relation being satisfied.
This result indicates that the AMM remains important
for any magnetic field strength and proton energy.

As written in Ref.~\cite{P2Pi-1}, the very large effect of the AMM comes
from the shift of this peak position, i.e. where the HO overlap integral
$\cM (n_i, n_f)$ changes rapidly.
For $n_i \approx 48$ and a shift of $\Delta n_{if}$ to 2, the absolute
value of $\cM$ increases by about a factor of 100.
When the magnetic field is weaker or the initial energy becomes larger,
$n_i$ increases, and the AMM effect is expected to be smaller.
When $\chi$ is fixed, however, the AMM effect remains and plays an
important role in any regime.

Furthermore, from Fig.~\ref{WidIFb},
we note that the Landau level difference in the pion emission
between the initial and final states is of the same order as that
of the initial and final Landau numbers,
$\Delta n_{if} \sim n_i \sim n_f$.

In the adiabatic limit it can be assumed that the relative momentum
between the final proton and the pion is zero, and that
the two particles have nearly the same velocity.
In that limit the ratio between  these two energies
is the same as the mass ratio: $e_\pi / e_f \approx m_\pi / m_p$,
and the final proton and pion energies become
\begin{equation}
e_f \approx \frac{m_p}{m_p + m_\pi} e_i, \quad
e_\pi \approx \frac{m_\pi}{m_p + m_\pi} e_i .
\end{equation}

If the initial proton energy is very large $e_i \gg m_p$,
$E_{i,f} \approx \sqrt{2 n_{i,f}}$, so that
in the adiabatic and high energy limit the following relation holds:
$ \sqrt{n_i} - \sqrt{n_f} \approx (m_\pi /m_p) \sqrt{n_i}$.
This leads to
\begin{eqnarray}
\Delta n_{if} \equiv n_i - n_f \approx
\frac{m_p^2 - (m_p -m_\pi)^2}{m_p^2} n_i
\approx 0.28 n_i .
\end{eqnarray}

The actual $\pi N$-interaction is via a $p$-wave, and the relative momentum
is not zero even in the adiabatic limit.
Thus, the actual value of  $\Delta n_{if}$ in Fig. 1 is larger than the above value;
this argument is consistent with the present results.

\begin{figure}[hbt]
\begin{center}
{\includegraphics[scale=0.6]{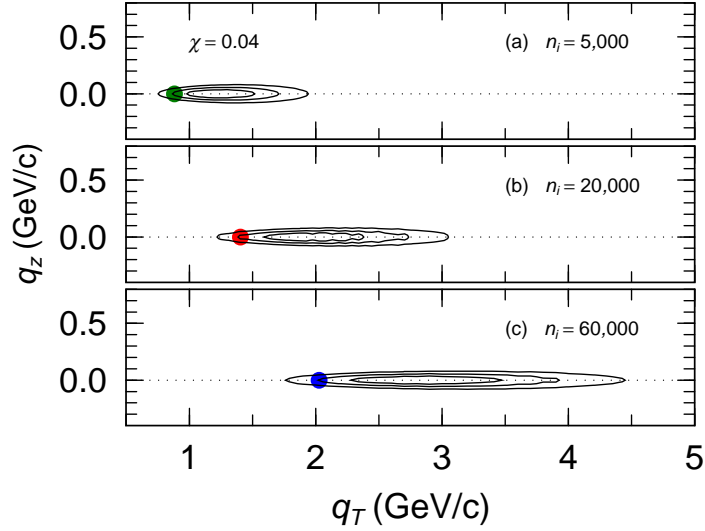}}
\caption{\small
(Color online)
Contour plot of the differential pion luminosity
for the initial spin state $s_i=-1$ and $\chi = 0.04$ at $p_{iz}$ = 0 
with initial Landau numbers,
 (a) $n_i = 5 \times 10^3$, (b) $2 \times 10^4$ and (c) $6 \times 10^4$.
Lines on each plot show the relative strength, 25\%, 50\% and 75\%.
The vertical and  horizontal axes are the $z$-component and the
transverse component of the emitted pion momentum.
The large dots show the adiabatic limit.
 }
\label{PiAng}
\end{center}
\end{figure}

In Fig.~\ref{PiAng}, we show contour plots of the luminosity distribution
of the emitted pions, $d^3 I / d q^3 = e_\pi d^3 \Gamma_{p \pi} / d q^3$,
where the initial proton is at $p_{iz} = 0$
and $n_i = 5 \times 10^3$ (a), $2 \times 10^4$ (b), or
$6 \times 10^4$ (c) with $\chi = 0.04$.
The pion momentum is distributed  narrowly in the $z$-direction
independently of $n_i$, but is broadly distributed
in the transverse direction.
The width in the $z$-direction is almost independent of $n_i$, but
the width in the transverse direction becomes larger
as the initial Landau number $n_i$ increases.  
It means that most pions are emitted in the transverse direction 
when $p_{iz}= 0$ in the limit of $n_i \rightarrow \infty$.

When $n_{i,f}$ and $s_{i,f}$ are fixed, the differential
decay width is proportional to the HO overlap integral $\cM (n_i, n_f)$
in Eq.~(\ref{TrStM}), which is an oscillating function with respect
to $Q_T^2$, $\cM (n_1,n_2) = \cM (n_1, n_2, Q_T^2)$, 
but the actual value of $Q_T^2$ in the present calculations turns out 
to be restricted to the region below the first peak.
When $Q_T^2 \ll 1$, $\cM (n_1, n_2, Q_T^2) \propto Q_T^{n_1 - n_2}$, and
thus $\cM$ is a monotonic and very rapidly increasing function in the region
for our calculations.
So, the decay width has an effective strength only
around the maximum value of $Q_T^2$. This effective region becomes narrower
as  $\Delta n_{if}$ increases.

The momentum region of the emitted pions when $p_{iz} \neq 0$ can then
be calculated from the results with $p_{iz} =0$
by a Lorentz transformation along the $z$-direction.
Since the pion is emitted in the transverse direction when $p_{iz} =0$,
the $z$-component of the velocity for the emitted pion and
the final proton are equal to that of the initial proton, namely
$P_{iz} / E_i = P_{fz} / E_f = Q_z / E_{\pi}$,
which leads to
$Q_z /Q_T \approx  P_{iz}/\sqrt{2 n_i}  \approx P_{iz}/ \sqrt{2 n_f}$.
Hence, when $p_{iz} \neq 0$,  the differential decay width is given by
\begin{equation}
\frac{d^3 \Gamma_{p \pi} (n_i, s_i)}{d q^3}
=  \frac{\sqrt{e_\pi^2 - q_z^2}}{2 \pi e^2_{\pi}} \sum_{n_f}
\Gamma (n_i, s_i, n_f, -s_i) \delta(e_\pi - e_i + e_f)
\delta \left( q_z - \frac{e_{\pi}}{e_i} p_{iz} \right) .
\label{ScDecay}
\end{equation}

\begin{wrapfigure}{r}{8.2cm}
\begin{center}
{\includegraphics[scale=0.42]{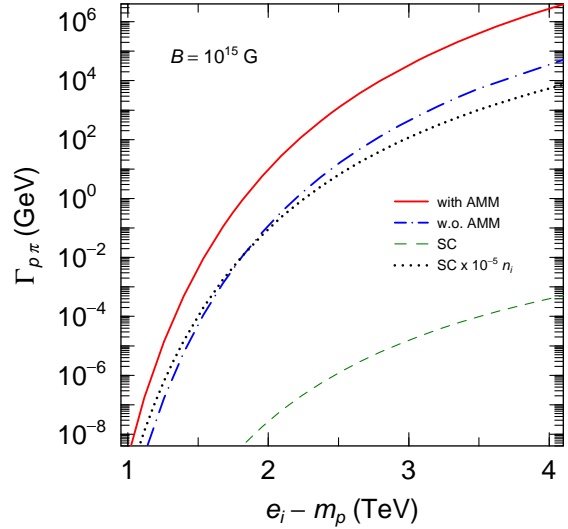}}
\caption{\small
(Color online)
Decay width of a proton with $p_{iz} = 0$ and $s_i = -1$
due to synchrotron emission
as a function of the proton initial energy $e_i - m_p$
when $B= 10^{15}$G.
The solid and dot-dashed lines represent the decay widths
 with and without the AMM,  respectively.
The dashed and dotted lines indicate the results in the semi-classical
 calculations \cite{TK99} and those multiplied with $10^{-5} n_i$,
respectively
}
\label{TWid}
\end{center}
\end{wrapfigure}

When the magnetic field is taken to be $B \sim 10^{15}$G, the initial energy of the proton should be a several TeV, and
the Landau number of the initial state should be order of $10^{12} - 10^{13}$.
Performing a quantum calculation is not realistic in such a condition.
However, we can calculate $\Gamma(n_i, n_f)$ with $n_i \sim 10^4$ and
extrapolate it to $n_i \sim 10^{12}$ via the above scaling relation.
In this way we can calculate the decay width in any realistic conditions.

Hereafter, we apply the scaling relation to the case of a realistic
strength of the magnetic field $B \sim 10^{15}$G
by using the above equation (\ref{ScDecay}).

In Fig.~\ref{TWid}  we show the pion decay widths for protons
with  $p_{iz} =0$ and $s_i = -1$ as functions of the initial energy
when  $B = 10^{15}$G.
The solid and dot-dashed lines represent the decay widths of the proton
with and without the AMM, respectively.
For comparison, we also give  the results in the semi-classical
 approaches of Ref.~\cite{TK99} with the dashed line.

In the quantum approach we calculate the decay width with
$n_i = 160,000 - 1,000,000$ and extrapolate them
to that when $n_i \sim 10^{12}$, by using the scaling relation.

First, we can confirm that the AMM still increases the decay widths
significantly
even in the realistic conditions where the magnetic fields are much weaker
and the initial Landau  numbers are much larger  than those
in the previous work, $B=5 \times 10^{18}$G and $n_i \approx 48$.
\cite{P2Pi-1}.

Second, we see that the quantum results are much larger than those
in the semi-classical approach.
As mentioned before, one assumes that $\Delta n_{if} \ll n_i$ in that
approach and that the phase-space of the final pion is very small.
In Ref.~\cite{TK99} they assumed that the pion energy is the same as the
pion mass $e_\pi = m_\pi$, so that
the total decay width  does not depend on the initial energy if  $\chi$
is fixed.

\begin{wrapfigure}{r}{8.5cm}
\begin{center}
\vspace{-01cm}
{\includegraphics[scale=0.55]{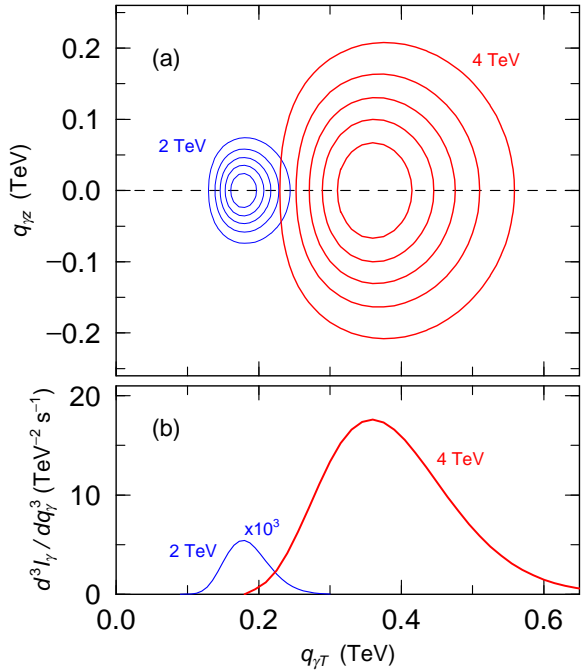}}
\caption{\small
(Color online)
The differential luminosities per a proton versus the photon momentum.
They are integrated over the initial proton angle.
Upper panel shows  contour plots at an initial proton energy
$e_i - m_p =2$ TeV (thin line) and 4 TeV (thick line).
Lower panel shows  the luminosity distribution  at $q_z = 0$ when $e_i - m_p = 2$~TeV
and $e_i - m_p  = 4$~TeV.
The former result is reduced by a factor of $10^4$.
 }
\label{PhMom}
\end{center}
\end{wrapfigure}

In the quantum calculation, on the other hand, the energy of the emitted
pion is of the same order as the initial proton energy,
and  the scaling relation about $\Gamma(n_i, n_f)$ shows
that the total decay width is proportional to the initial landau number,
$\Gamma_{p \pi} (n_i) \propto n_i$.

In order to examine this, we multiply by a factor
proportional to the initial Landau number, $10^{-5} n_i$,
with the decay width in the semi-classical calculation
and show this result with the dotted line.
This result is qualitatively similar to that without the AMM in the quantum
approach; we should note that if we use the factor, $10^{-3} n_i$, 
the result is close to that with the AMM.
It can be conjectured from this result that the main difference
in the total decay width between the quantum approach and
the semi-classical approach comes from the naive estimate of the energy of
the emitted pions and their phase space volume.

Next, we study the  photon luminosity distribution,
$d^3 I_\gamma  /d q_\gamma^3$ at $q_{\gamma z} = 0$,
for a realistic magnetic field strength of $B = 10^{15}$G,
where $q_\gamma$ is  the momentum of the  emitted photon which is 
a half of the emitted pion momentum, $q_\gamma = q/2$, 
in the ultra-relativistic limit.

We assume that the momentum distribution of the initial proton
is spherical and make an average of the luminosity over the proton angles.
In Fig.~4 we show the luminosity distribution per a proton
when the initial proton energies are $e_i - m_p = 2$ and 4 TeV,
and the magnetic field $B = 10^{15}$G;
this is a more realistic representation of the magnetar environment.

In the upper panel (a) the (blue) thin and (red) thick lines show
 contour plots for $e_i - 2m_p = 2$~TeV and 4 TeV,
respectively.
In the lower panel (b) we show the distribution $d^3 I_\gamma  /d q^3$ at $q_z = 0$ for
$e_i - m_p = 2$~TeV and 4~TeV.

As noted above, the  incident proton emits a pion, whose energy is
$\sim 10-30$ \% (depending upon $\chi$) of the incident proton energy, 
along the same direction as that of the incident proton momentum.
This emitted pion decays into two photons with the same energy and the
same direction of their movement in this ultra-relativistic energy region.

The proton transverse momentum is proportional to $\sqrt{\chi}$, and the
decay width rapidly decreases as  $\chi$ become smaller.
Then, the luminosities mainly distribute in a region perpendicular to the
magnetic field, though they also distribute about 40 - 60 \% of the photon momentum along the direction of the magnetic
field.

In summary, we have calculated the exact pion decay width of protons
in the relativistic quantum approach including the Landau levels
and the AMM for a magnetic field strength in the range
$B = 10^{18} - 10^{17}$~G, where the maximum Landau levels are
$n_{max} \approx 5 \times 10^3 - 10^5$.
As the magnetic field becomes weaker, and the Landau number
tremendously increases, 
the polar angular distributions of the emitted pion momentum becomes narrower,
when the initial and final proton Landau numbers are fixed.
Then, the polar angles of the emitted pion and the final proton momenta
are almost the same as that of the initial proton.

The large effect of the AMM and the very narrow width of the polar
angular distribution are caused by a rapid change of the HO
overlap integral $\cM (n_i, n_f)$.

As the energy of incident proton increases or the magnetic field is
weaker, 
the distribution of emitted pion momentum $|\vq|$ on the $q_z - q_T$
plane is elongated in the radial direction, 
keeping the width along $q_z$  narrow, 
because the dominant transition between  the two Landau levels is
as large as  
$\Delta n_{if} / n_i \gtrsim 0.3$ in  $\Gamma (n_i, n_f)$.
As such, the quantity  $\Gamma (n_i, n_f)$ 
depends only on $\Delta n_{if}/n_i$ and $\chi =eB e_i / m_p^3$.

In the usual semi-classical approximations it is assumed that
$\Delta n_{if} \equiv n_i - n_f \ll n_i$,
where the HO overlap integral $\cM$ can be approximated 
with an Airy function.
This assumption is equivalent to the adiabatic limit
when the produced particle is massless such as a photon.

Furthermore, the very low energy of emitted particles leads to a small phase
space volume in the  final particle momentum space and causes  the total
decay width to be underestimated.
This underestimation becomes larger as the initial proton energy
increases.
As the mass of the produced particle becomes larger, its emitted momentum
increases.
Thus, our calculation suggests a change in the semi-classical relation
between the production rate and $\chi$.

In conclusion, the present work suggests a better way
to treat pion production.
Although direct calculation with a realistic magnetic field and
a large number of Landau levels is not tractable,
the result that $\Gamma (n_i, n_f)$ depends only on $\chi$ and
$\Delta n_{if}/n_i$ demonstrates that one can calculate  $\Gamma (n_i, n_f)$
for values of $n_i \sim 10^{4-5}$
and scale those results to more realistic conditions.
Also, when $p_{iz} \neq 0$, the decay width can be obtained
from a Lorenz transformation along the $z$-direction.

By using this scaling relation we can, for the first time, present
the luminosity distribution
of  photons due to the pion production process in  realistic environments 
where the magnetic field is $B = 10^{15}$G and the incident proton
energy is $e_i - m_p = 2$ and 4 TeV.

In this work, we have found  the energy distribution of emitted particles and
the large effect of the AMM in the particle production
from the synchrotron radiation. 
These features turn out to be caused by properties of the HO overlap function,
$\cM (n_i, n_f)$, which is written in terms of the associated Laguerre function,
$L_{n_f}^{n_i - n_f} (Q_T^2 / 2)$  \cite{P2Pi-1}. 
We do not know its asymptotic form when $n_i \sim n_f \rightarrow
\infty$, and we cannot prove the above features analytically at present, 
although we deduce an efficient scaling relation in the function.
If we knew it, we could make a general formulation for all kinds of
particle production.
This we leave to a future work.

In other future works, we will be able to calculate the emitted photon 
distribution
and the magnetic structure of  magnetars using our method
and obtain significant information from observations of energetic  
photons from magnetars.

\medskip

This work was supported in part by Grants-in-Aid for
Scientific Research of JSPS (26105517, 24340060)
of the Ministry of Education, Culture, Sports, Science
and Technology of Japan,
and also by the National Research Foundation of Korea
(Grants No. NRF-2014R1A2A2A05003548).
Work at the University of Notre Dame is supported
by the U.S. Department of Energy under
Nuclear Theory Grant DE-FG02-95-ER40934.


\end{document}